# Quasi-Periodic WaveNet Vocoder: A Pitch Dependent Dilated Convolution Model for Parametric Speech Generation

*Yi-Chiao Wu[1], Tomoki Hayashi[2], Patrick Lumban Tobing[1], Kazuhiro Kobayashi[3], Tomoki Toda[3]*

[1] Graduate School of Informatics, Nagoya University, Japan
[2] Graduate School of Information Science, Nagoya University, Japan
[3] Information Technology Center, Nagoya University, Japan
yichiao.wu@g.sp.m.is.nagoya-u.c.jp, tomoki@icts.nagoya-u.c.jp

## Abstract

In this paper, we propose a quasi-periodic neural network (QPNet) vocoder with a novel network architecture named pitch-dependent dilated convolution (PDCNN) to improve the pitch controllability of WaveNet (WN) vocoder. The effectiveness of the WN vocoder to generate high-fidelity speech samples from given acoustic features has been proved recently. However, because of the fixed dilated convolution and generic network architecture, the WN vocoder hardly generates speech with given $F_0$ values which are outside the range observed in training data. Consequently, the WN vocoder lacks the pitch controllability which is one of the essential capabilities of conventional vocoders. To address this limitation, we propose the PDCNN component which has the time-variant adaptive dilation size related to the given $F_0$ values and a cascade network structure of the QPNet vocoder to generate quasi-periodic signals such as speech. Both objective and subjective tests are conducted, and the experimental results demonstrate the better pitch controllability of the QPNet vocoder compared to the same and double-size WN vocoders while attaining comparable speech qualities.

## 1. Introduction

For conventional parametric speech synthesis, speech is usually decomposed into several acoustic features and synthesized with these acoustic features. The analysis-synthesis technique is called a vocoder [1], and the foundation of a vocoder is a speech production mechanism based on source excitation and vocal tract. The main advantage of a vocoder is that it provides high flexibility for users to manipulate the synthesized speech to meet their scenarios. However, because of the oversimplified assumptions from conventional vocoders such as STRAIGHT [2] and WORLD [3], temporal details and phase information are lost, and it causes significant quality degradation.

Recently, neural network (NN) based speech synthesis [4–11] has become one of the most popular techniques, which is widely applied to many devices in daily life such as speech assistants and car navigators. However, human perception is quite sensitive to speech quality, and that of synthesized speech highly depends on the generation model. WaveNet (WN) [4] is one of the state-of-the-art speech generation models, which has been applied to many applications, such as speech enhancement [12, 13], text-to-speech (TTS) [7, 9], speech coding [11], and voice conversion (VC) [15–18]. Specifically, WN is an autoregressive model that predicts a current speech sample based on a specific number of previous samples which is called the *receptive field*. Because of the long-term dependence of speech signals, WN applies a stacked dilated convolution network (DCNN) structure to efficiently extend the *receptive field*. Furthermore, non-autoregressive generation models [19, 20] also have been proposed to reduce the generation time while maintaining the comparable speech qualities to WN. For NN-based vocoders, the WaveNet vocoder [21–23], which is a WaveNet conditioned on the acoustic features extracted by a traditional vocoder to generate speech, achieves significant improvements in speech naturalness than traditional vocoders.

However, it is difficult for the WN vocoder to deal with unseen conditional features. That is, the WN vocoder cannot generate relevant speech from the given fundamental frequency ($F_0$) that outside the range observed in training data, whereas the pitch controllability is an essential mechanism of traditional vocoders. This difficulty may be caused by the fixed network architectures. Specifically, the fixed *receptive field* indicates that each speech sample correlates to the same numbers of past samples, but it is more reasonable that each sample has its own dependent field. Furthermore, to generate high-fidelity speech, the required long *receptive field* results in a huge network size.

To tackle these problems, we propose a quasi-periodic NN-based (QPNet) vocoder with a novel pitch-dependent dilated convolution network (PDCNN), which is inspired by source-filtering model [24] and code-excited linear prediction (CELP) codec [25], to model the relationships of speech samples in a pitch cycle with the short-term correlation and then extend that to whole quasi-periodic signals with the long-term correlation. Specifically, QPNet includes two cascaded WNs with different dilated convolution structures. The first part is the original WN using the fixed DCNNs to presumably generate signals based on a specific segment of previous samples. The second part including PDCNNs makes the network generate signals based on the relevant segments of previous cycles. The adaptive dilated structure deals with the unseen $F_0$ with the introduced quasi-periodic information and gives each sample an exclusive *receptive field* corresponding to the conditional $F_0$. Moreover, because the proposed QPNet vocoder extneded the *receptive field* more efficiently than the original WN vocoder, half the network size was required to achieve acceptable performance according to the experimental results.

## 2. WaveNet vocoder

The WN vocoder models the long-term dependence among sequential waveform samples and auxiliary acoustic features using a conditional probability as follows:

$$P(\mathbf{Y} | \mathbf{h}) = \prod_{t=1}^{T} P(y_t | y_{t-1}, ..., y_{t-r}, \mathbf{h}),  \qquad (1)$$

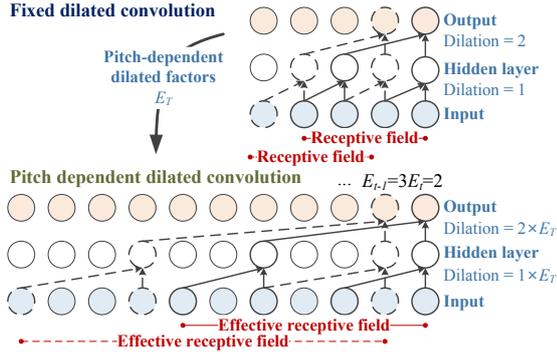
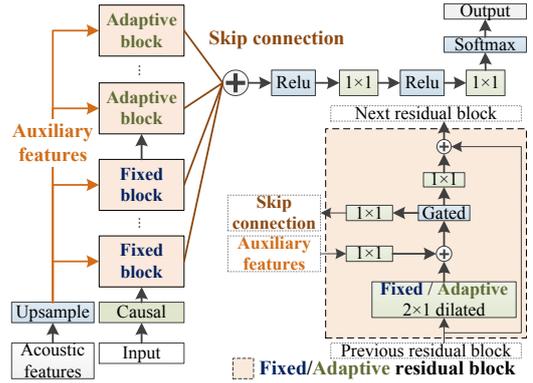

Figure 1: *Pitch-dependent dilated convolution*

Figure 2: *Quasi-Periodic WaveNet vocoder architecture*

where $t$ is the sample index, $r$ is the length of the *receptive field*, $y_t$ is the current audio sample, and $\mathbf{h}$ is the auxiliary feature vector. That is, the WN vocoder predicts the conditional distribution of the current speech sample with input auxiliary features and a specific number of previous samples, which is called the *receptive field*. Furthermore, the WN vocoder usually transforms the speech generation into a classification problem. By encoding speech signals into 8 bits using the $\mu$-law, the output of the WN vocoder becomes a categorical distribution. In addition, a gated structure is applied to enhance the modeling capability which is formulated as

$$\mathbf{Z} = \tanh\left(\mathbf{V}_{f,k}^{(1)} * \mathbf{Y} + \mathbf{V}_{f,k}^{(2)} * u(\mathbf{h})\right) \otimes \sigma\left(\mathbf{V}_{g,k}^{(1)} * \mathbf{Y} + \mathbf{V}_{g,k}^{(2)} * u(\mathbf{h})\right), \quad (2)$$

where $\mathbf{V}^{(1)}$ and $\mathbf{V}^{(2)}$ are trainable convolution filters, $*$ is the convolution operator, $\otimes$ is an elementwise multiplication operator, $\sigma$ is a sigmoid function, $k$ is the layer index, $f$ and $g$ represent the filter and gate, respectively, and $u(\cdot)$ is an upsampling layer used to adjust the resolution of auxiliary features to match that of input speech samples. Moreover, because of the very long term dependence and causality of speech signals, WaveNet applies a DCNN structure [4, 26] to guarantee the causality and efficiently extend the *receptive field*. To sum up, previous speech samples pass through a pipeline including a causal layer and several residual blocks which contain a dilated convolution layer, gated activation with auxiliary features, and residual and skip connections. Then, the summation of all skip connections is passes to two 1×1 convolution and one softmax layers to output the predicted distribution of the current sample.

However, because of the data-driven nature without the speech related prior knowledge, the WN vocoder lacks the pitch controllability. For example, the traditional vocoders based on source-filtering model easily generate speech with precise pitches matched to arbitrarily input $F_0$ values, but the WN vocoder often has the difficulty in generating speech or tends to generate speech within the $F_0$ range observed in training data when conditioned on the unseen $F_0$ values.

## 3. Quasi-Periodic WaveNet vocoder

The cascaded structure of the autoregressive networks and the pitch-dependent mechanism of the dilated convolution neural networks of QPNet are inspired by the short/long-term prediction architectures and the pitch filtering technique of CELP. The details are as follows.

### 3.1. Pitch filtering in CELP

For CELP, a given excitation sequence from a codebook is filtered by a linear-prediction and pitch filters to reconstruct speech. The linear-prediction filter restores the spectral (short-term correlation) information. The pitch (long-delay) filter generates the pitch periodicity of the voiced speech follows

$$c_o[t] = G \times c_i[t] + b \times c_o[t - t_d], \quad (3)$$

where $c_i[t]$ is the input, $c_o[t]$ is the output, $G$ is the gain, $b$ is the pitch filter coefficient and $t_d$ is the pitch delay.

### 3.2. Pitch-dependent dilated convolution

Figure 1 elaborates the concept of PDCNN. If the input is a sequential quasi-periodic signal with time-variant $F_0$, the *receptive field* lengths of the original structure (fixed dilated convolution) are time-invariant but that of the pitch-dependent one are changed corresponding to the $F_0$ values. Specifically, the dilated convolution is formulated as

$$\mathbf{X}_t^{(o)} = \mathbf{W}^{(c)} * \mathbf{X}_t^{(i)} + \mathbf{W}^{(p)} * \mathbf{X}_{t-d}^{(i)}, \quad (4)$$

where $\mathbf{X}^{(i)}$ and $\mathbf{X}^{(o)}$ are the input and output of the DCNN layer. $\mathbf{W}^{(c)}$ and $\mathbf{W}^{(p)}$ are the trainable 1×1 convolution filters of current and past samples, respectively. The dilation size $d$ is a constant for DCNN but time-variant for PDCNN. Specifically, the pitch-dependent $d$ makes the *receptive field* of each sample with arbitrary pitch contain a specific number of previous cycles. That is, the network predicts each current sample given the same number of previous cycles, while each current sample has the different pitch. Therefore, the pitch-dependent structure makes the network efficiently extend the *receptive field* without losing trajectory information of the sequential signals.

In addition, for original stacked DCNN, the dilation size is doubled for every layer up to a specific number and then repeated. Proposed PDCNN also follows the same rule to layer-wise extend the dilation sizes but with an extra dilated factor $E_t$ to adjust the dilation sizes to match the pitch of the current sample. The pitch-dependent dilated factor $E_t$ is as follows:

$$E_t = F_s / (F_{0,t} \times a), \quad (5)$$

where $F_s$ is the sampling rate which is a constant of the whole utterance, $F_{0,t}$ is the fundamental frequency of speech sample with sample index $t$, and $a$ is a hyperparameter. Therefore, each speech sample has a pitch-matched length of the *receptive field*. Furthermore, $a$ indicates the number of samples in one cycle for considering, and we empirically set it to 8 in this paper. We also applied the interpolated continuous $F_0$ rather than the discrete ones to get the pitch-dependent dilated factors because of the better performance based on our internal experiments.

Table 1: *Comparison of hyperparameters*

| Hyperparameters | WNf | WNc | QPNet |
|---|---|---|---|
| Number of fixed layers | 10 | 4 | 4 |
| Number of fixed repeats | 3 | 4 | 3 |
| Number of adaptive layers | - | - | 4 |
| Number of adaptive repeats | - | - | 1 |
| Constant $a$ | - | - | 8 |
| Causal and dilated conv. | 512 channels | | |
| 1×1 conv. in residual blocks | 512 channels | | |
| 1×1 conv. between skip-connection and softmax | 256 channels | | |

Table 2: *Comparison of root-mean-square error of log $F_0$ with different vocoders and $F_0$ transformed ratios*

| $F_0$ ratio | WORLD | WNf | WNc | QPNet |
|---|---|---|---|---|
| Unchanged | 0.09 | 0.14 | 0.26 | **0.13** |
| 1/2 | 0.13 | 0.30 | 0.38 | **0.23** |
| 2/3 | 0.11 | 0.23 | 0.35 | **0.19** |
| 3/4 | 0.10 | 0.20 | 0.32 | **0.17** |
| 4/5 | 0.10 | 0.18 | 0.30 | **0.16** |
| 6/5 | 0.09 | 0.16 | 0.26 | **0.13** |
| 5/4 | 0.09 | 0.17 | 0.26 | **0.14** |
| 4/3 | 0.10 | 0.18 | 0.26 | **0.15** |
| 3/2 | 0.09 | 0.21 | 0.27 | **0.16** |
| 2 | 0.09 | 0.26 | 0.28 | **0.18** |
| Average | 0.10 | 0.20 | 0.29 | **0.16** |

Table 3: *Comparison of Mel-cepstral distortion with different vocoders and $F_0$ transformed ratios*

| $F_0$ ratio | WORLD | WNf | WNc | QPNet |
|---|---|---|---|---|
| Unchanged | 2.52 | **3.58** | 4.34 | 4.08 |
| 1/2 | 3.92 | **4.56** | 5.02 | 4.79 |
| 2/3 | 3.19 | **4.15** | 4.71 | 4.47 |
| 3/4 | 2.93 | **3.95** | 4.58 | 4.34 |
| 4/5 | 2.79 | **3.84** | 4.50 | 4.27 |
| 6/5 | 2.72 | **3.60** | 4.38 | 4.14 |
| 5/4 | 2.76 | **3.62** | 4.39 | 4.16 |
| 4/3 | 2.83 | **3.63** | 4.42 | 4.19 |
| 3/2 | 3.05 | **3.68** | 4.50 | 4.27 |
| 2 | 3.75 | **3.86** | 4.75 | 4.59 |
| Average | 3.04 | **3.84** | 4.56 | 4.33 |

### 3.3. Cascaded autoregressive networks

Figure 2 shows the architecture of the proposed QPNet vocoder that consists of two main modules. The first module is like the original WN vocoder to have a causal layer and several stacked residual blocks including a dilated convolution, conditional auxiliary features, gated activations, and residual and skip connections. The second module also has several stacked adaptive residual blocks similarly to the first module but alternatively adopting the pitch-dependent dilated convolutions. Furthermore, motivated by CELP, we cascade the two modules to respectively model the short and long-term dependences of speech signals. Specifically, based on the assumption that speech can be decomposed into periodic and nonperiodic components, we assume that the nonperiodic parts mostly depend on the nearest samples, while the periodic parts have very long term dependences. Therefore, the first module of QPNet is used to estimate the short-term information, and the second module models the long-term periodical correlations.

## 4. Experiments

### 4.1. Experimental settings

We conducted objective and subjective tests to evaluate the performance of four vocoders including the WORLD [3] vocoder, the WN vocoders with two different network sizes, and the proposed QPNet vocoder. Specifically, we trained a compact-size QPNet vocoder to compare with a compact-size WaveNet (WNc), full-size WaveNet (WNf), and WORLD vocoders. The hyperparameters of the networks are sown in Table 1. The training followed our previous work [16].

The training corpus of the multispeaker WNf, WNc, and QPNet vocoders included the training data of the 'bdl' and 'slt' speakers of CMU-ARCTIC [27] and all training data of VCC2018 [28] consistented with [16]. The four source speakers (two males and two females) of the SPOKE set of VCC2018 were used as an evaluation set, and each speaker contained 35 testing utterances. The original acoustic features were extracted by WORLD, which consisted of one-dimensional $F_0$ and 513-dimensional spectral (*sp*) and aperiodic (*ap*) features. $F_0$ was converted into continuous $F_0$ features and voice/unvoice (*uv*) binary symbols, *sp* was parameterized into 34-dimensional Mel-cepstrum coefficient (*mcep*), and *ap* were coded into two-dimensional components [18]. Furthermore, we simulated outside unseen acoustic features by scaling $F_0$ with ten different ratios from 1/2 to 2. The following evaluations were conducted on the basis of the transformed $F_0$, original natural *mcep*, coded *ap*, WORLD vocoder, and multispeaker vocoders of WNc, WNf, and QPNet.

### 4.2. Objective evaluations

For the objective tests, we respectively measured the pitch accuracy and spectral distortion of the generated speech using the root-mean-square error (RMSE) of logarithmic $F_0$ and Mel-cepstral distortion (MCD). Specifically, to evaluate the pitch generation accuracy of each vocoder related to the conditional $F_0$, we calculated the RMSE between the conditional $F_0$ and the $F_0$ extracted from the generated speech. Moreover, we computed MCD between the conditional and extracted *mcep* to evaluate the robustness of the spectrum reconstruction, while conditioned on the unseen acoustic features.

Table 2 shows the results of RMSE of log $F_0$, and we can find that the proposed QPNet vocoder significantly outperformed the same-size WNc vocoder. Even compared to the double-size WNf vocoder, the QPNet vocoder still achieved better pitch generation accuracy, especially conditioned on the scaled $F_0$ with large offset. Although the conventional WORLD vocoder reasonably achieved the lowest RMSE, the proposed QPNet vocoder still remarkably improved the accuracy of pitch generation with unseen conditional $F_0$ compared to the original WN vocoders. In addition, Table 3 indicates that the proposed QPNet vocoder still had much better spectrum prediction capability than the same size WNc vocoder. However, the QPNet vocoder got worse performance than the WNf vocoder.

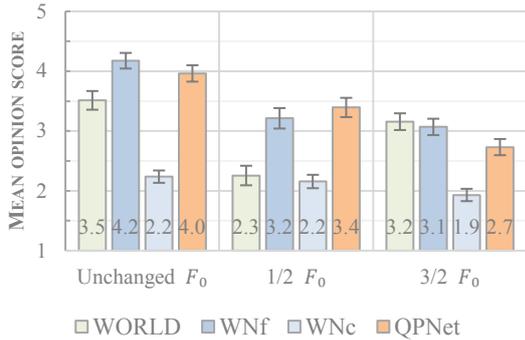

Figure 3: *MOS evaluation of sound quality with 95% confidence intervals.*

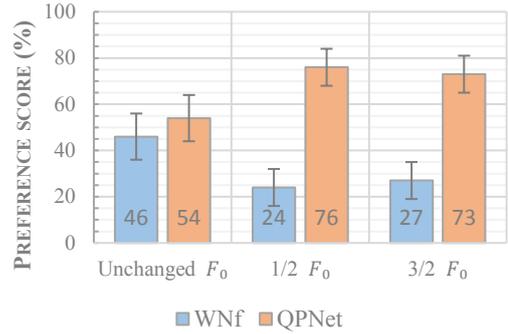

Figure 4: *XAB evaluation of pitch accuracy with 95% confidence intervals.*

The much shorter *receptive field* length caused by the halved network size might degrade the spectral prediction capability of QPNet. In summary, the objective evaluations confirm that the proposed pitch-dependent dilation structure can improve the pitch generation accuracy for NN-based vocoders.

### 4.3. Subjective evaluations

For the perceptual evaluations, we conducted the mean opinion score (MOS) and XAB preference tests to evaluate the sound quality and pitch accuracy of the generated utterances from different vocoders conditioned on the acoustic features with different scaled $F_0$. Specifically, we randomly selected 20 utterances from 35 testing utterances of each speaker and scaled $F_0$ to form an evaluation set. Then, we divided the set into five subsets and each one was evaluated by two subjects. The total number of subjects was 10, and the demo can be found at "https://bigpon.github.io/QuasiPeriodicWaveNet_demo/".

For the MOS test, the subjects evaluated the 960 utterances which were generated using the WORLD, WNf, WNc, and QPNet vocoders given the acoustic features with unchanged, 1/2, and 3/2 $F_0$. The measurements were 1~5 and the higher score meant the better sound quality. For the XAB test, each subject first listened to one reference and two testing utterances and then selected the testing utterance that had more consistent pitches with the reference one. Moreover, because we did not have the real speech with scaled $F_0$, and the conventional vocoders could generate speech with more precise pitches in the unseen $F_0$ scenarios, we took the WORLD generated utterances as the reference speech. That is, the subjects evaluated the pitch accuracy of the WNf and QPNet generated utterances based on the WORLD generated reference utterances.

As shown in Fig. 3, in the inside $F_0$ range (unchanged $F_0$) case, although WORLD achieved better MCD, WNf still got much better MOS. The oversimplified excitation model of WORLD caused serious buzz noise, and WNf generated speech without many handcraft assumptions and achieved better perceptual quality. However, this result also indicates that the performance of the WN vocoder highly depends on the length of *receptive field*, so the quality of the WNc generated speech significantly degraded. As a result, after applying PDCNN, which can efficiently extend the *receptive field*, the QPNet vocoder achieved comparable sound qualities to the WNf vocoder. Moreover, Fig. 4 suggests that QPNet achieved comparable pitch generation accuracy with WNf, which is consistent with the objective results. In addition, in the outside 1/2 $F_0$ cases, WORLD suffered severe naturalness degradation especially in very low $F_0$ (male speakers) cases which made WORLD generate robotic speech. However, conditioned on 1/2 $F_0$, QPNet and WNf still generated speech with acceptable quality as shown in Fig. 3, and QPNet got remarkably higher pitch generation accuracy as shown in Fig. 4. In the outside 3/2 $F_0$ cases, WORLD showed the robustness against arbitrary $F_0$ inputs. Although QPNet still attained higher pitch accuracy, the sound quality of QPNet became worse than that of WNf.

### 4.4. Discussion

We selected a compact network size which was only half of the original WN vocoder, so only about 75 % training and 40% generation times were required. However, it made the *receptive field* length become much shorter. For example, the length of the *receptive field* of WNf was 3070 (The *receptive field* length of 10 layers in each repeat was $2^0+2^1+ \ldots 2^9=1023$, so the total length was 1023×3 with extra one from causal layer), but that of WNc was only 61 ($2^0+2^1+2^2+2^3=15$ in each repeat, and the total *receptive field* length was 15×4+1=61). Furthermore, the effective *receptive field* length of QPNet was 46+15*$E_t$ (The *receptive field* length of fixed and causal layers was 15×3+1=46, and that of the adaptive layers was the product of 15 and the pitch-dependent dilated factors), so the size was around 886 to 136 for the $F_0$ range of training corpus was around 50 to 500 Hz with sampling rate of 22.05 kHz (the pitch-dependent dilated factor of 50 Hz was 56 that was the ceiling of 22050/(50×8)), which were quite shorter than that of WNf. It is the possible reason that QPNet achieved worse speech quality while the auxiliary $F_0$ with high scaled ratio.

## 5. Conclusions

In this paper, we proposed a QPNet vocoder with the new pitch-dependent dilated convolution which extends the *receptive field* more efficiently than the WN vocoder. Moreover, the QPNet vocoder also has higher pitch generation accuracy, which takes advantage of the proposed PDCNN, and comparable sound quality to the double-size WN vocoder. In conclusion, the QPNet vocoder is more in line with the definition of vocoder. In future works, we will survey the effectiveness of QPNet in the voice conversion task.

## 6. Acknowledgements

This work was partly supported by JST, PRESTO Grant Number JPMJPR1657, and JSPS KAKENHI Grant Number 17H01763.